\documentclass[print,twocolumn,showpacs,aps]{revtex4}%
\usepackage{mathrsfs}
\usepackage{amsfonts}
\usepackage{amsmath}
\usepackage{amssymb}
\usepackage{graphicx}%
\begin{document}
\title{Symmetry breaking and manipulation of nonlinear optical modes in an asymmetric
double-channel waveguide}
\author{Rujiang Li$^{1}$, Fei Lv$^{1}$, Lu Li$^{1}$ }
\email{llz@sxu.edu.cn}
\author{Zhiyong Xu$^{2}$}
\affiliation{$^{1}$Institute of Theoretical Physics, Shanxi University, Taiyuan 030006, China}
\affiliation{$^{2}$Nonlinear Physics Centre, Research School of Physics and Engineering,
The Australian National University, Canberra ACT, 0200, Australia}

\begin{abstract}
We study light beam propagation in a nonlinear coupler with an asymmetric
double-channel waveguide, and derive various analytical forms of optical
modes. The results show that the symmetry-preserving modes in a symmetric
double-channel waveguide is deformed due to the asymmetry of the two-channel
waveguide, and such a coupler supports yet the symmetry-breaking modes. The
dispersion relations reveal that the system with self-focusing nonlinear
response supports the degenerate modes, while for self-defocusing medium the
degenerate modes do not exist. Furthermore, nonlinear manipulation is
investigated by launching optical modes supported in double-channel waveguide
into a nonlinear uniform medium.

\end{abstract}
\maketitle

\section{introduction}

Propagation of optical waves in waveguide arrays has become an important and
effective means to investigate various optical phenomena which have analogues
in many fields of physics \cite{Longhi}. Special attention has been paid to
nonlinear surface waves and nonlinear guided waves in planar layered
structures \cite{M1,M2,M3,M4,M5,M6,M7,M8,M9,M10,M11,M12} and nonlinear
couplers \cite{Marcuse,Yangcc,Yangcc1,Chen,Yasumoto}, generation and
properties of solitons in nonlinear waveguide arrays
\cite{Nat,Lederer,PO,YS,Ablowitz,YS4} and other nonlinear periodic systems,
such as optically induced lattices \cite{MS,Neshev,Yaroslov,Zhangyp}. The
investigation of light beam propagation in waveguide arrays attracts
increasing attention due to their potential applications in all-optical signal
processing in fiber optic networks and devices, the passive mode-locking by
using waveguide array \cite{R2}, and the beam steering \cite{PRE53,ZhangOC,Xu}.

The behavior of light beam propagation in a coupler composed by two-channel
nonlinear waveguide gained particular attention because it can exhibit some of
universal properties in nonlinear periodic systems and nonlinear waveguide
arrays \cite{Yaroslov1}. It is shown that the coupler can support the
symmetry-preserving solutions which have the linear counterparts and the
symmetry-breaking solutions without any linear counterparts
\cite{Kevrekidis,Birnbaum,Jia}, in which the spontaneous symmetry-breaking has
been experimentally demonstrated in optically induced lattices with a local
double-well potential \cite{Kevrekidis}.

In this paper, we consider light beam propagation in an asymmetric
double-channel waveguide with Kerr-type nonlinear response, and derive various
analytical stationary solutions in detail. It is found that the asymmetric
double-channel waveguide can break the symmetric form of the
symmetry-preserving modes otherwise in the symmetric double-channel waveguide,
and such a coupler supports the symmetry-breaking modes. We also investigate
how the type of nonlinear response affects the existence and properties of
nonlinear optical modes in the asymmetric double-channel waveguide. The
dispersion relation shows that the degenerate modes exist in the system with
self-focusing nonlinear response, while for the coupler with self-defocusing
response the degenerate modes do not exist. In addition, based on these
optical modes supported in asymmetric double-channel waveguide we demonstrate
the control and manipulation of optical modes in different nonlinear media by
tuning the waveguide parameters.

The paper is organized as follows. In Section II, the model equation
describing beam propagation in a double-channel waveguide is derived. In
Section III, various analytical forms of optical modes are presented both in
self-focusing and self-defocusing media. Meanwhile, the dispersion relations
between the total energy and the propagation constant are discussed. In
Section IV, we study the nonlinear manipulation of optical modes in
double-channel waveguide. The conclusion is summarized in Section V.

\section{Model equation and reductions}

We consider a planar graded-index waveguide with refractive index%
\begin{equation}
n(z,x)=F(x)+n_{2}I(z,x). \label{ref_ind}%
\end{equation}
Here the first term on the right hand side presents a two-channel waveguide
with the different refractive index, namely, $F(x)=$ $n_{11}$ as
$-L_{0}/2-D_{0}<x<-L_{0}/2$, and $F(x)=n_{12}$ as $L_{0}/2<x<L_{0}/2+D_{0}$,
otherwise, $F(x)=$ $n_{0}$ ($<n_{11},n_{12}$), where $D_{0}$ and $L_{0}$
represent the width of waveguide and the separation between waveguides,
respectively; while $n_{0}$, $n_{11}$ and $n_{12}$ denote the refractive index
of cladding and waveguide, respectively. The second term denotes Kerr-type
nonlinearity, $I(z,x)$ is the optical intensity, and positive (negative) value
of nonlinear coefficient $n_{2}$ indicates self-focusing (self-defocusing)
medium. Under slowly varying envelope approximation, the nonlinear wave
equation governing beam propagation in such a waveguide with the refractive
index given by Eq. (\ref{ref_ind}) can be written as
\begin{equation}
i\frac{\partial\psi}{\partial z}+\frac{1}{2k_{0}}\frac{\partial^{2}\psi
}{\partial x^{2}}+\frac{k_{0}\left[  F(x)-n_{0}\right]  }{n_{0}}\psi
+\frac{k_{0}n_{2}}{n_{0}}\left\vert \psi\right\vert ^{2}\psi=0, \label{eq1}%
\end{equation}
where $\psi(z,x)$ is the envelope function, and $k_{0}=2\pi n_{0}/\lambda$ is
wave number with $\lambda$ being wavelength of the optical source generating
the beam. By introducing the normalized transformations $\psi(z,x)=(k_{0}%
\left\vert n_{2}\right\vert L_{D}/n_{0})^{-1/2}\varphi(\zeta,\xi)$,
$\xi=x/w_{0}$ and $\zeta=z/L_{D}$ with $L_{D}=2k_{0}w_{0}^{2}$, which
represents the diffraction length, we get the dimensionless form of Eq.
(\ref{eq1}) as follows%
\begin{equation}
i\frac{\partial\varphi}{\partial\zeta}+\frac{\partial^{2}\varphi}{\partial
\xi^{2}}+V(\xi)\varphi+\eta\left\vert \varphi\right\vert ^{2}\varphi=0,
\label{eq2}%
\end{equation}
where $\eta=n_{2}/\left\vert n_{2}\right\vert =\pm1$ corresponds to
self-focusing ($+$) and self-defocusing ($-$) nonlinearity of the waveguides,
respectively, and $V(\xi)=2k_{0}^{2}w_{0}^{2}\left[  F(w_{0}\xi)-n_{0}\right]
/n_{0}$ is of the form
\begin{equation}
V(\xi)=\left\{
\begin{array}
[c]{cc}%
V_{1}, & -L/2-D<\xi<-L/2,\\
V_{2}, & L/2<\xi<L/2+D,\\
0, & \text{otherwise,}%
\end{array}
\right.  \label{V}%
\end{equation}
which describes the dimensionless two-channel waveguide structure with
different refractive index, where $V_{1}=2k_{0}^{2}w_{0}^{2}(n_{11}%
-n_{0})/n_{0}$ and $V_{2}=$ $2k_{0}^{2}w_{0}^{2}(n_{12}-n_{0})/n_{0}$ being
the modulation depth of the refractive index of the left and right waveguide,
and $L=L_{0}/w_{0}$ and $D=D_{0}/w_{0}$ corresponding to scaled separation and
width of waveguide, respectively. Here, we use the typical waveguide
parameters $D=3.5$, $L=5$, $V_{2}=2.525$, and vary $V_{1}$. Figure 1 shows the
profile of the two-channel waveguide structure given by Eq. (\ref{V}). It
should be pointed out that such structure can be realized experimentally
\cite{FL}. It is shown that the Eq. (\ref{eq2}) conserves the total energy
flow $P(\zeta)=\int_{-\infty}^{+\infty}\left\vert \varphi(\zeta,\xi
)\right\vert ^{2}d\xi=P_{0}$, where $P_{0}$ is the dimensionless initial total energy.

\begin{figure}[ptb]
\centering\vspace{-0.0cm} \includegraphics[width=6.0cm]{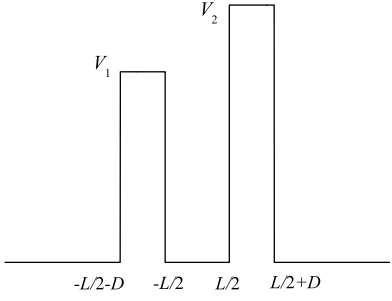}
\vspace{-0.3cm}\caption{The profile of a two-channel waveguide with different
refractive index.}%
\end{figure}

Assuming that the stationary solution of Eq. (\ref{eq2}) is of the form
$\varphi(\zeta,\xi)=u(\xi)\exp(i\beta\zeta)$, where $u(\xi)$ is a real
function, and $\beta$ is the propagation constant, and substituting it into
Eq. (\ref{eq2}), we find that the function $u(\xi)$ obeys the following
nonlinear equation%
\begin{equation}
\frac{d^{2}u}{d\xi^{2}}+\left[  V(\xi)-\beta\right]  u+\eta u^{3}=0,
\label{sta_equ}%
\end{equation}
where $\eta=\pm1$ corresponds to self-focusing ($+$) and self-defocusing ($-$)
nonlinearity of the waveguides, respectively. It should be pointed out that
Eq. (\ref{sta_equ}) not only can describe the optical modes in the
double-channel waveguide structure with the different refractive index, but
also can describe one-dimensional Bose-Einstein condensate trapped in a finite
asymmetry double square well potential $-V(\xi)$. In particular, when
$V_{1}=V_{2}$, the corresponding optical modes in the symmetric double-channel
waveguide structure have been studied, and the results have shown that the
coupler not only supports symmetry-preserving modes but also symmetry-breaking
modes \cite{Jia}.

\section{Optical modes}

In this section, we will present the analytical solutions of Eq.
(\ref{sta_equ}) with the potential (\ref{V}) for $\eta=\pm1$. Generally, the
solutions of Eq. (\ref{sta_equ}) can be constructed in terms of the Jacobi
elliptic functions depending on the values of the variable $\xi$, and have the
same propagation constants in different regions. Within the double-channel
waveguides of $-L/2-D<\xi<-L/2$ and $L/2<\xi<L/2+D$, the solution of Eq.
(\ref{sta_equ}) is the oscillatory function, so it can be selected in the form
\cite{WDLi}%
\begin{equation}
u_{1}(\xi;A,K,\delta)=A\operatorname*{sn}\left(  K\xi+\delta,-\frac{\eta
A^{2}}{2K^{2}}\right)  , \label{sn}%
\end{equation}
with $\beta=V_{1}-K^{2}+\eta A^{2}/2$ in the region of $-L/2-D<\xi<-L/2$ and
$\beta=V_{2}-K^{2}+\eta A^{2}/2$ in the region of $L/2<\xi<L/2+D$. In the
region of $\left\vert \xi\right\vert <L/2$, the solution of Eq. (\ref{sta_equ}%
) has two different Jacobi elliptic functions for the symmetric and the
antisymmetric case, respectively. For the symmetric case, the solution is
\cite{WDLi}
\begin{equation}
u_{2}(\xi;B,Q,\sigma)=B\operatorname*{nc}\left(  Q\xi+\sigma,1+\frac{\eta
B^{2}}{2Q^{2}}\right)  , \label{nc}%
\end{equation}
with $\beta=Q^{2}+\eta B^{2}$; and for the antisymmetric case, the solution is
\cite{WDLi}%
\begin{equation}
u_{2}(\xi;B,Q,\sigma)=B\operatorname*{sc}\left(  Q\xi+\sigma,1+\frac{\eta
B^{2}}{2Q^{2}}\right)  , \label{sc}%
\end{equation}
with $\beta=Q^{2}-\eta B^{2}/2$. It should be noted that those two solutions
are precise solutions of Eq. (\ref{sta_equ}) for one node and no node within
the region of $\left\vert \xi\right\vert <L/2$. In other regions, the solution
of Eq. (\ref{sta_equ}) is required to tend to zero as $\xi\rightarrow\pm
\infty$, so it is taken as \cite{Jia}
\begin{equation}
u_{3}(\xi;b)=\frac{1}{be^{-\sqrt{\beta}\xi}+ce^{\sqrt{\beta}\xi}},
\label{sech}%
\end{equation}
with $\beta>0$. Substituting (\ref{sech}) into Eq. (\ref{sta_equ}), one
obtains $c=\eta/(8\beta b)$.

Note here that although the modulus in the usual Jacobi elliptic function is
restricted from 0 to 1, this problem can be solved by the modular
transformation such that the modulus in the Jacobi elliptic functions given by
Eqs. (\ref{sn}-\ref{sc}) can take any positive or negative values in our
investigations, as shown in Refs. \cite{Byrd,WDLi}, so those solutions do not
depend on nonlinearity sign.

In the following, we show the analytical global solutions of Eq.
(\ref{sta_equ}). With the help of Eqs. (\ref{sn}), (\ref{nc}) [or (\ref{sc})],
and (\ref{sech}), the solutions of Eq. (\ref{sta_equ}) can be written as%
\begin{equation}
u(\xi)=\left\{
\begin{array}
[c]{ll}%
u_{3}(\xi;b_{1}), & \text{ \ }\xi<-L/2-D,\\
u_{1}(\xi;A_{1},K_{1},\delta_{1}), & \text{ \ }-L/2-D<\xi<-L/2,\\
u_{2}(\xi;B,Q,\sigma), & \text{ \ }\left\vert \xi\right\vert <L/2,\\
u_{1}(\xi;A_{2},K_{2},\delta_{2}), & \text{ \ }L/2<\xi<L/2+D,\\
u_{3}(\xi;b_{2}), & \text{ \ }\xi>L/2+D.
\end{array}
\right.  \label{solution}%
\end{equation}
The continuity conditions for $u$ and $\partial u/\partial\xi$ at the
boundaries of $\xi=\pm L/2$ and $\xi=\pm(L/2+D)$ require%
\begin{align}
u_{3}\left(  -L/2-D;b_{1}\right)   &  =u_{1}\left(  -L/2-D;A_{1},K_{1}%
,\delta_{1}\right)  ,\nonumber\\
\frac{du_{3}}{d\xi}\left(  -L/2-D;b_{1}\right)   &  =\frac{du_{1}}{d\xi
}\left(  -L/2-D;A_{1},K_{1},\delta_{1}\right)  ,\nonumber\\
u_{1}\left(  -L/2;A_{1},K_{1},\delta_{1}\right)   &  =u_{2}(-L/2;B,Q,\sigma
),\nonumber\\
\frac{du_{1}}{d\xi}\left(  -L/2;A_{1},K_{1},\delta_{1}\right)   &
=\frac{du_{2}}{d\xi}(-L/2;B,Q,\sigma),\nonumber\\
u_{2}(L/2;B,Q,\sigma)  &  =u_{1}(L/2;A_{2},K_{2},\delta_{2}),\nonumber\\
\frac{du_{2}}{d\xi}(L/2;B,Q,\sigma)  &  =\frac{du_{1}}{d\xi}(L/2;A_{2}%
,K_{2},\delta_{2}),\nonumber\\
u_{1}(L/2+D;A_{2},K_{2},\delta_{2})  &  =u_{3}(L/2+D;b_{2}),\nonumber\\
\frac{du_{1}}{d\xi}(L/2+D;A_{2},K_{2},\delta_{2})  &  =\frac{du_{3}}{d\xi
}(L/2+D;b_{2}), \label{con}%
\end{align}
with $\beta=V_{1}-K_{1}^{2}+\eta A_{1}^{2}/2=V_{2}-K_{2}^{2}+\eta A_{2}^{2}%
/2$, and $\beta=Q^{2}+\eta B^{2}$ for the symmetric case given by Eq.
(\ref{nc}) or $\beta=Q^{2}-\eta B^{2}/2$ for the antisymmetric case given by
Eq. (\ref{sc}). In Eq. (\ref{solution}), there are eleven parameters $A_{1}$,
$K_{1}$, $\delta_{1}$, $A_{2}$, $K_{2}$, $\delta_{2}$, $B$, $Q$, $\sigma$,
$b_{1}$, and $b_{2}$, which can be calculated by solving numerically Eqs.
(\ref{con}) with the conditions that the propagation constants in different
regions should be same. Once those parameters are determined numerically, we
can obtain the exact optical modes for asymmetric double-channel nonlinear waveguides.

\begin{figure}[ptb]
\centering\vspace{-0cm} \includegraphics[width=9cm]{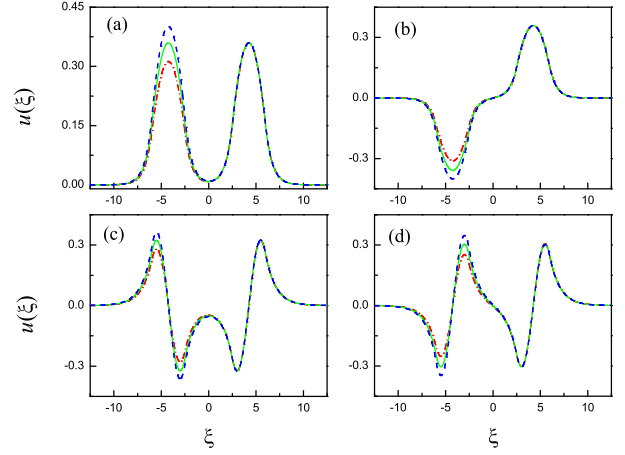}
\vspace{-0.5cm}\caption{(Color Online) Various different optical modes
existing in self-defocusing medium ($\eta=-1$), where the dash-dotted red line
is the optical mode for $V_{1}=2.500$, the solid green line is the optical
mode for $V_{1}=2.525$, and the dashed blue line is the optical mode for
$V_{1}=2.550$. Here $\beta=2.00$ in (a) and (b), $\beta=0.85$ in (c) and (d).}%
\end{figure}

\begin{figure}[ptb]
\centering\vspace{-0cm} \includegraphics[width=9cm]{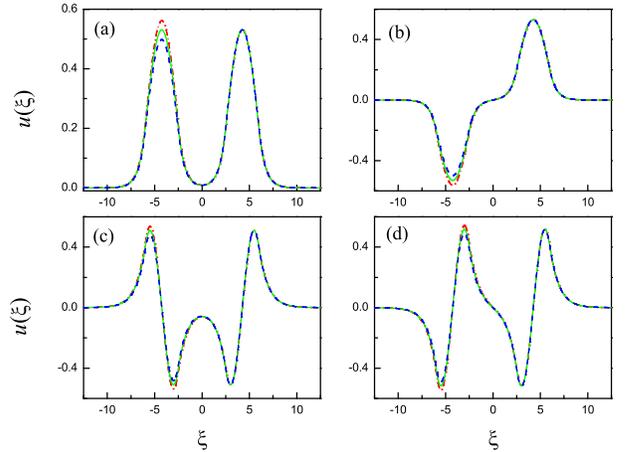}
\vspace{-0.5cm}\caption{(Color Online) Various different optical modes
existing in self-focusing medium ($\eta=1$), where the dash-dotted red line is
the optical mode for $V_{1}=2.500$, the solid green line is the optical mode
for $V_{1}=2.525$, and the dashed blue line is the optical mode for
$V_{1}=2.550$. Here $\beta=2.30$ in (a) and (b), $\beta=1.10$ in (c) and (d).}%
\end{figure}

Fig. 2 and Fig. 3 show several different optical modes in a nonlinear
asymmetric double-channel waveguide in the self-defocusing medium and the
self-focusing medium, respectively. These optical modes are induced from the
symmetry-preserving optical modes in the symmetric double-channel waveguide,
where for comparison, we also plotted the corresponding symmetry-preserving
optical modes in the symmetric double-channel waveguides in the same figure.
From Figs. 2 and 3, we found that the symmetry of the modes is broken due to
asymmetry of the two-channel waveguide, and the amplitude of the modes in the
lower refractive index waveguide is smaller than that in the higher refractive
index waveguide for the self-defocusing medium, while for the self-focusing
medium, the amplitude of the modes in the lower refractive index waveguide is
larger than that in the higher refractive index waveguide.

\begin{figure}[ptb]
\centering\vspace{-0cm} \includegraphics[width=9cm]{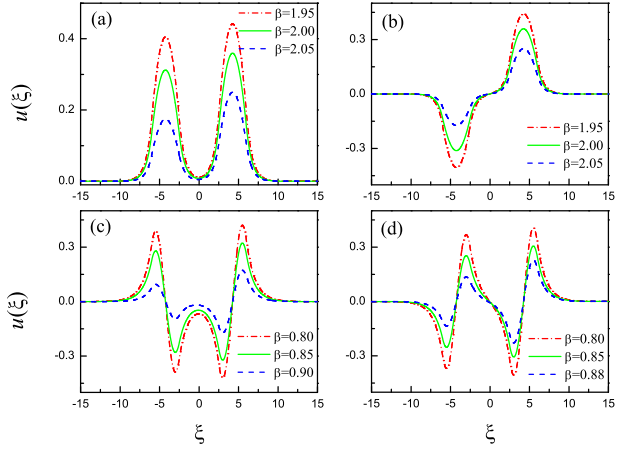}
\vspace{-0.5cm}\caption{(Color Online) Several optical modes with different
$\beta$ in a nonlinear asymmetrical double-channel waveguide for the
self-defocusing medium ($\eta=-1$). Here the parameters are $V_{1}=2.500$ and
$V_{2}=2.525$.}%
\end{figure}

\begin{figure}[ptb]
\centering\vspace{-0cm} \includegraphics[width=9cm]{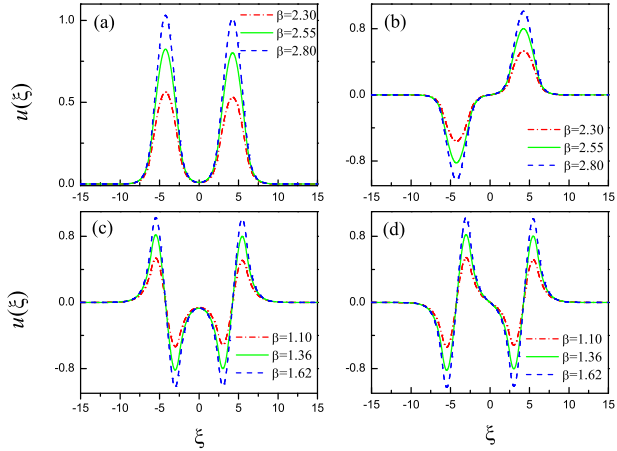}
\vspace{-0.5cm}\caption{(Color Online) Several optical modes with different
$\beta$ in a nonlinear asymmetrical double-channel waveguide for the
self-focusing medium ($\eta=1$). Here the parameters are the same as in Fig.
4.}%
\end{figure}

We also demonstrate the profiles of optical modes with dependence on the
propagation constant $\beta$. Fig. 4 and Fig. 5 present several corresponding
modes shown in Figs. 2 and 3 for different propagation constant $\beta$. From
them, one can see that, for self-defocusing media the profile of nonlinear
modes is shrunk, and the corresponding amplitude becomes smaller with an
increase of the propagation constant $\beta$, while for self-focusing case it
is opposite, namely, the profile of nonlinear modes becomes more prominent and
the corresponding amplitude becomes larger. This feature can be depicted by
the dispersion relations between the total energy $P_{0}$ and the propagation
constant $\beta$. As shown in Fig. 8 and Fig. 9, one can see that the total
energy decreases with the increase of the propagation constant for
self-defocusing media (see Fig. 8), whereas it is an increasing function of
the propagation constant for self-focusing case (see Fig. 9).

\begin{figure}[ptb]
\centering\vspace{-0cm} \includegraphics[width=9cm]{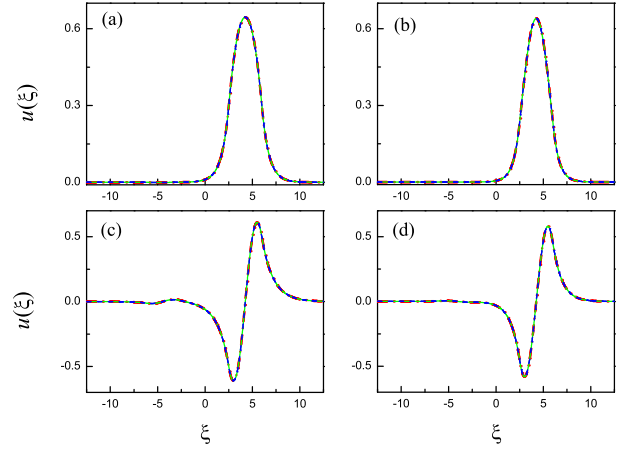}
\vspace{-0.5cm}\caption{(Color Online) Several symmetry-breaking optical
modes, where the dash-dotted red lines are optical modes for $V_{1}=2.500$,
the solid green lines are optical modes for $V_{1}=2.525$, and the dashed blue
lines are optical modes for $V_{1}=2.550$. Here $\eta=-1$ in (a) and (c),
$\eta=1$ in (b) and (d) and $\beta=1.78$ in (a), $\beta=2.39$ in (b),
$\beta=0.66$ in (c) and $\beta=1.15$ in (d).}%
\end{figure}

\begin{figure}[ptb]
\centering\vspace{-0cm} \includegraphics[width=9cm]{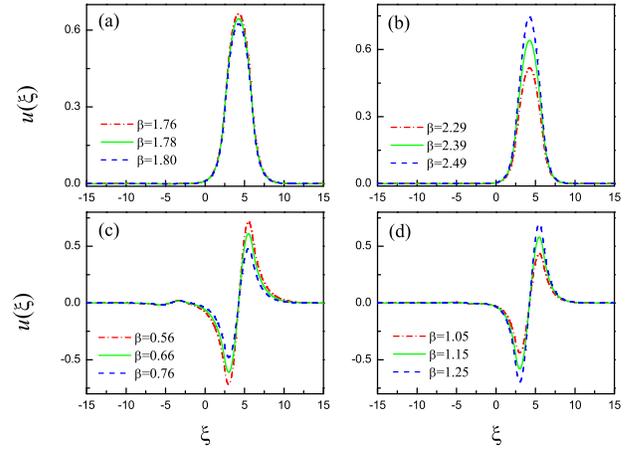}
\vspace{-0.5cm}\caption{(Color Online) Several symmetry-breaking optical modes
with different $\beta$ in a nonlinear asymmetrical double-channel waveguide.
Here the parameters are the same as in Fig. 6.}%
\end{figure}

As discussed in Ref. \cite{Jia}, besides the symmetry-preserving optical
modes, a double-channel waveguide also supports the symmetry-breaking optical
modes, and the corresponding optical modes in a nonlinear asymmetric
double-channel waveguide are presented in Fig. 6, in which we also plotted the
corresponding symmetry-breaking optical modes in a symmetric double-channel
waveguide in the same figure for comparison. One can find that the optical
modes in a nonlinear asymmetric double-channel waveguide are almost the same
as the modes in a symmetric one for the given $\beta$.

Similarly, the corresponding modes shown in Fig. 6 for different propagation
constant $\beta$ are presented in Fig. 7. It is shown that the profile of
nonlinear modes becomes shrunk with the increase of the propagation constant
$\beta$ for the self-defocusing case, while for the self-focusing case the
profile of nonlinear modes becomes more pronounced. This feature is depicted
by the dispersion relations between the total energy $P_{0}$ and the
propagation constant $\beta$. It should be pointed out that the modes shown in
Fig. 6 only exist in a small region, as shown in Fig. 8 and Fig. 9.

\begin{figure}[ptb]
\centering\vspace{-0cm} \includegraphics[width=11cm]{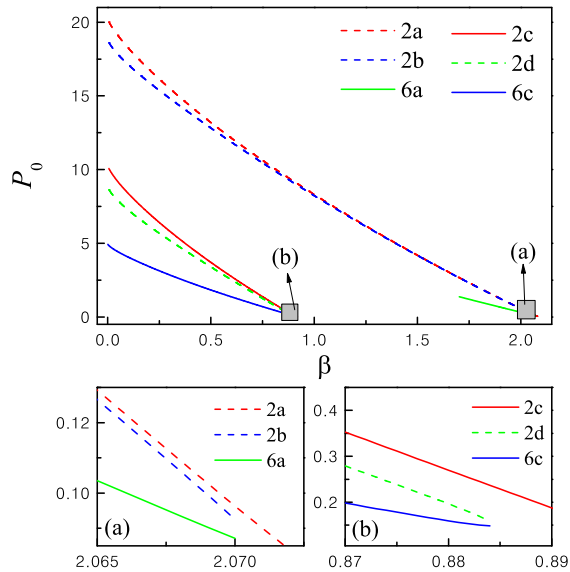}
\vspace{-0.5cm}\caption{(Color Online) The dependence of the total energy
$P_{0}$ on the propagation constant $\beta$ for the modes existing in
self-defocusing medium ($\eta=-1$). Here the parameters are $V_{1}=2.500$ and
$V_{2}=2.525$. The labeled shadow areas are enlarged in corresponding (a)-(b).
And the labels 2a, 2b, $\cdots$, mean the corresponding modes shown in Fig.
2a, 2b, $\cdots$, respectively.}%
\end{figure}

\begin{figure}[ptb]
\centering\vspace{-0cm} \includegraphics[width=11cm]{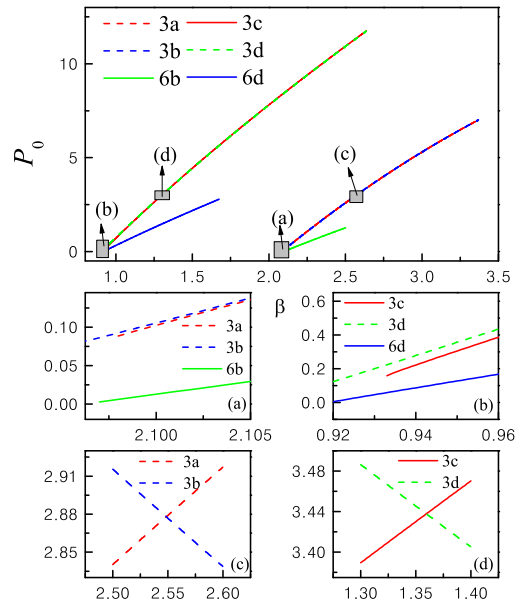}
\vspace{-0.5cm}\caption{(Color Online) The dependence of the total energy
$P_{0}$ on the propagation constant $\beta$ for the modes existing in
self-focusing medium ($\eta=1$). Here the parameters are $V_{1}=2.500$ and
$V_{2}=2.525$. The labeled shadow areas are enlarged in corresponding (a)-(d).
And the labels 3a, 3b, $\cdots$, mean the corresponding modes shown in Fig.
3a, 3b, $\cdots$, respectively.}%
\end{figure}

From the dispersion relations shown in Figs. 8 and 9, one can see that for the
self-defocusing medium there is no intersection between dispersion curves (see
Fig. 8 and the enlarged Figs. 8a and 8b), which indicates that there is no
degenerate modes existing, and the total energy of the mode shown in Fig. 2a
is the highest for a given propagation constant $\beta$. While for the
self-focusing medium the dispersion curves can intersect (see the enlarged
Figs. 9c and 9d), which implies that there exists two different modes with the
same total energy at the intersection point, namely, the degeneracy occurs at
the intersection point. Note that for our choices of the parameters, the
intersection points of the dispersion curves for the modes shown in Figs. 3a
and 3b, and Figs. 3c and 3d are about $2.5485$ and $1.3595$, respectively, and
the corresponding degenerate modes are shown by the green solid curves in Fig.
5. Here, to distinguish the case of the intersection, we rotate the dispersion
curve for the modes shown in Fig. 3a (Fig. 3c) for an angle of $\pi/6$
anticlockwise as the center of intersection point and the same angle for
dispersion curve of the mode shown in Fig. 3b (Fig. 3d) but rotate clockwise,
as shown in Figs. 9c and 9d.

\begin{figure}[ptb]
\includegraphics[width=9.5cm]{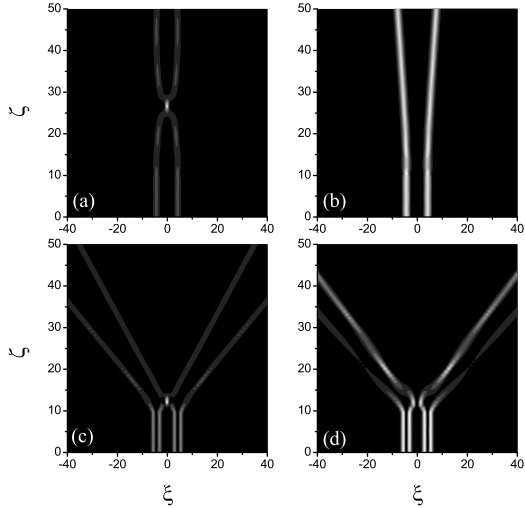} \vspace{0.0cm}\caption{(Color
Online) The evolution of optical modes shown in Fig. 2 into the self-focusing
Kerr medium without channels, where $\eta^{\prime}=10$ in (a) and (b), and
$\eta^{\prime}=20$ in (c) and (d). Here the parameters are $V_{1}=V_{2}=2.525$
and $\zeta_{0}=10$. The labels (a), (b), (c), and (d) mean the corresponding
modes shown in Fig. 2a, 2b, 2c, and 2d, respectively.}%
\end{figure}

\begin{figure}[ptb]
\includegraphics[width=9.5cm]{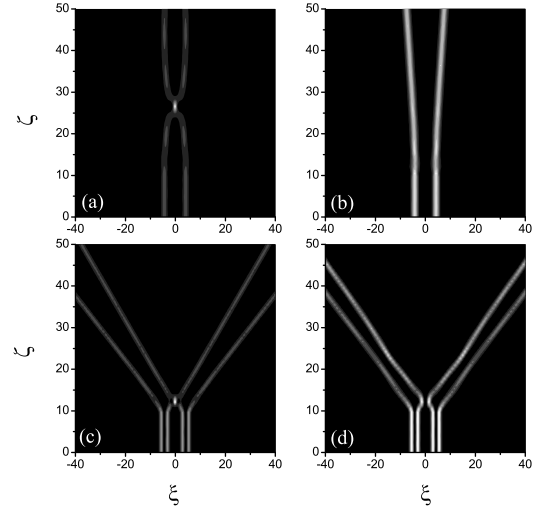} \vspace{0.0cm}\caption{(Color
Online) The evolution of optical modes shown in Fig. 3 into the self-focusing
Kerr medium without channels, where $\eta^{\prime}=5$ in (a) and (b), and
$\eta^{\prime}=10$ in (c) and (d). Here the parameters are $V_{1}=V_{2}=2.525$
and $\zeta_{0}=10$. The labels (a), (b), (c), and (d) mean the corresponding
modes shown in Fig. 3a, 3b, 3c, and 3d, respectively.}%
\end{figure}

\section{The nonlinear manipulation of optical modes in double-channel
waveguide}

In this section, we will demonstrate the control and manipulation of optical
modes in reconfigurable nonlinear media. Thus our interest is to investigate
the evolution dynamics of optical beams existing in double-channel waveguide
propagating into a uniform nonlinear medium. In this case, the governing
equation can be generally written as%
\begin{equation}
i\frac{\partial\psi}{\partial z}+\frac{1}{2k_{0}}\frac{\partial^{2}\psi
}{\partial x^{2}}+\frac{k_{0}\Delta n(z,x)}{n_{0}}\psi=0, \label{Gen_eq}%
\end{equation}
where the refractive index change $\Delta n(z,x)$ is a function of $z$ and
$x$, and $\Delta n(z,x)=n(z,x)-n_{0}$. Here we assume that when $0\leq z\leq
Z_{0}$, $n(z,x)$ is in the form of Eq. (\ref{ref_ind}), $\Delta
n(z,x)=F(x)+n_{2}I(z,x)-n_{0}$; while for $z>Z_{0}$, $\Delta n(z,x)=n_{2}%
^{\prime}I(z,x)-n_{0}$. Here, $n_{2}$ and $n_{2}^{\prime}$ are the Kerr
nonlinear coefficients of different media in the region of $0\leq z\leq Z_{0}$
and $z>Z_{0}$, respectively. Thus, when $0\leq z\leq Z_{0}$, namely in the
region of $0\leq\zeta\leq\zeta_{0}$, Eq. (\ref{Gen_eq}) can be normalized to
Eq. (\ref{eq2}), where $\zeta=z/L_{D}$ and $\zeta_{0}=Z_{0}/L_{D}$ with
$L_{D}=2k_{0}w_{0}^{2}$. At the same time, in the region of $z>Z_{0}$, namely
$\zeta>\zeta_{0}$, Eq. (\ref{Gen_eq}) is reduced to the dimensionless form as
follows%
\begin{equation}
i\frac{\partial\varphi}{\partial\zeta}+\frac{\partial^{2}\varphi}{\partial
\xi^{2}}+\eta^{\prime}\left\vert \varphi\right\vert ^{2}\varphi=0,
\label{eq22}%
\end{equation}
where $\eta^{\prime}=n_{2}^{\prime}/\left\vert n_{2}\right\vert $. Note that
Eq. (\ref{eq22}) is different from Eq. (\ref{eq2}), in which Eq. (\ref{eq2})
includes a potential function $V(\xi)$ given by Eq. (\ref{V}), while Eq.
(\ref{eq22}) does not include and can describe the dynamics of beams in Kerr
media without any refractive index modulation.

In the following analysis, optical beams of different modes existing in
double-channel waveguide are injected into the uniform nonlinear medium after
propagating diffraction length of $\zeta_{0}$ in double-channel waveguide.

First we consider the situation that optical beams are injected from
symmetrical double-channel waveguide. For the optical modes shown in Fig. 2,
which exist in self-defocusing medium ($\eta=-1$), the numerical simulations
show that when $\eta^{\prime}<0$, these optical modes are diffracted quickly
after entering into a uniform Kerr medium. However, when $\eta^{\prime}>0$ and
is large enough, the evolution of optical beams exhibits different scenarios
in the Kerr medium without any channels, as shown in Fig. 10. Similarly, for
the optical modes shown in Fig. 3, which exist in self-focusing media
($\eta=1$) with double-channel waveguide, as shown in Fig. 11, our numerical
simulations show that the evolution of optical modes exhibit almost similar
properties with that in Fig. 10.

From Figs. 10 and 11, one can see that when the optical modes existing both in
self-defocusing and self-focusing media with double-channel waveguide are
injected into the self-defocusing medium without any channels, the beams
should be diffracted quickly. However,\ when the optical modes are injected
into the self-focusing medium without any channels and the corresponding
nonlinear coefficient $\eta^{\prime}$ is enough larger, the beams could be
manipulated effectively. In this situation, when the optical modes shown in
Fig. 2a and Fig. 3a are injected into self-focusing medium without any
channels the beams appear to attract and repel each other periodically, as
shown in Figs. 10a and 11a. While when the modes shown in Figs. 2b-2d and
Figs. 3b-3d are injected into self-focusing medium without any channels, the
beams repel each other, as shown in Figs. 10b-10d and Figs. 11b-11d. Note that
the escape angle are the same for the beams in Fig. 2b (Fig. 3b) due to the
symmetry of the double-channel waveguide.

\begin{figure}[ptb]
\includegraphics[width=9.5cm]{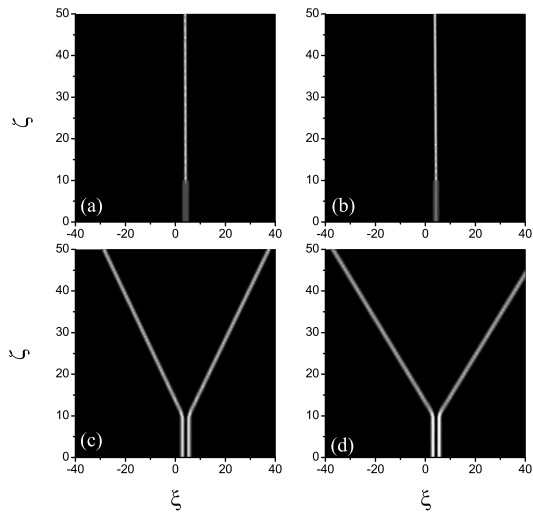} \vspace{-0.5cm}\caption{(Color
Online) The evolution of optical modes shown in Fig. 6 into the self-focusing
Kerr medium without channels, where $\eta^{\prime}=10$. Here the parameters
are $V_{1}=V_{2}=2.525$ and $\zeta_{0}=10$. The labels (a), (b), (c), and (d)
mean the corresponding modes shown in Fig. 6a, 6b, 6c, and 6d, respectively.}%
\end{figure}

\begin{figure}[t]
\includegraphics[width=9cm]{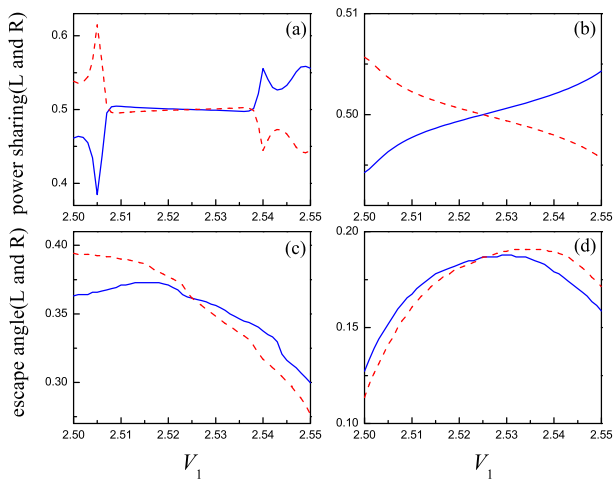} \vspace{-0.5cm}\caption{(Color
Online) Energy sharing (row 1) and escape angles (row 2) of optical beams as a
function of parameter $V_{1}$. In all cases, the solid-blue and the dashed-red
curves correspond to the left and right beams, respectively. Here the
parameter are $V_{2}=2.525$, $\eta^{\prime}=10$ in (a) and (c) corresponding
to the modes shown in Fig. 2b (namely Fig. 11b) and $\eta^{\prime}=5$ in (b)
and (d) corresponding to the modes shown in Fig. 3b (namely Fig. 12b).}%
\end{figure}

The evolution of optical modes shown in Fig. 6 is presented in Fig. 12, in
which Figs. 12a and 12c (Figs. 12b and 12d) demonstrate the evolution of
optical modes in the self-focusing medium without any channels with initial
input beams injected from self-defocusing (self-focusing) medium with
double-channel waveguide. As shown in Fig. 12a and 12b, one can see that
optical beams with a single hump can be compressed effectively. and as shown
in Figs. 12c and 12d the optical modes with two peaks are separated during the
evolution due to the repulsive interaction force resulted from the phase
difference between the two peaks.

It should be pointed out that these results only take into account the optical
modes existing in the symmetrical double-channel waveguide. Then, one will
naturally ask what is the influence of the asymmetrical double-channel on the
evolution of optical modes. In order to understand this question, we launch
optical beams from an asymmetrical double-channel waveguide into the
self-focusing medium to observe their evolution by tuning the depth of left
channel of the waveguide for a fixed depth of the right channel, namely, the
value of $V_{1}$ varies from $2.500$ to $2.550$ for $V_{2}=2.525$. As an
example, we demonstrated the evolution dynamics for the modes shown in Fig. 2b
and Fig. 3b. In Fig. 13, we present the dependence of the energy sharing, the
ratio of the energy carried by each component in the mode over the total
energy, and the escape angle, the angle of the each peak in the mode and the
propagation direction $\zeta$, on the value of $V_{1}$. As shown in Fig. 13,
one can see that the energy carried by each beam is different due to the
asymmetry of the double-channel waveguide [shown in Figs. 13a and 13b]. Also,
one can see clearly that the escape angles of the two beams take different
values with the change of the value $V_{1}$, which means that the beams can be
controlled by tuning the depth of the left channel of the waveguide.

\section{Conclusion}

We have studied light beam propagation in an asymmetric double-channel
waveguide in the form of a nonlinear coupler. A family of analytical solutions
with symmetric and antisymmetric forms has been obtained for both
self-focusing and self-defocusing nonlinear media, and the dispersion
relations between the total energy and the propagation constant has been
discussed in detail. Our results reveal that the system with self-focusing
nonlinear response supports the degenerate modes, while for self-defocusing
medium the degenerate modes do not exist. In addition, we explored new ways to
steer optical modes propagating from double-channel waveguide into a uniform
self-focusing medium. The compression of beam with single hump and split of
beams with two humps have been demonstrated by tuning the depth of the channel
of the waveguide. These properties may be applied in practical optical
devices, and be useful for optical processing, optical switching or optical routing.

\section{\textbf{ACKNOWLEDGEMENT}}

The authors acknowledge fruitful discussions with Professor Yuri Kivshar. This
research is supported by the National Natural Science Foundation of China
grant 61078079, the Shanxi Scholarship Council of China grant 2011-010, and
the Australian Research Council.

\end{document}